\documentclass[journal]{IEEEtran}
\usepackage{epsfig}
\usepackage{indentfirst}
\usepackage{amsmath}
\usepackage{amssymb}
\usepackage{times}
\usepackage{subfigure}
\usepackage{cite}
\usepackage{textcomp}
\usepackage{float}
\usepackage{color}
\usepackage{framed}
\usepackage{listings}

\makeatletter
\def\ps@headings{%
\def\@oddhead{\mbox{}\scriptsize\rightmark \hfil \thepage}%
\def\@evenhead{\scriptsize\thepage \hfil \leftmark\mbox{}}%
\def\@oddfoot{}%
\def\@evenfoot{}}
\makeatother
\IEEEoverridecommandlockouts
\begin{document}

\title{IoT Stream Processing and Analytics in The Fog}
\author{\IEEEauthorblockN{$\text{Shusen Yang}$}\\
\IEEEauthorblockA{National Engineering Laboratory for Big Data Algorithms and Analytics Technology,\\ Xi'an Jiaotong University, China}\\
}

\maketitle

\begin{abstract}
The emerging Fog paradigm has been attracting increasing interests from both academia and industry, due to the low-latency, resilient, and cost-effective services it can provide. 

Many Fog applications such as video mining and event monitoring, rely on data stream processing and analytics, which are very popular in the Cloud, but have not been comprehensively  investigated in the context of Fog architecture.
In this article, we present the general models and architecture of Fog data streaming, by analyzing the common properties of several typical applications.
We also analyze the design space of Fog streaming with the consideration of four essential dimensions (system, data, human, and optimization), where both new design challenges and the issues arise from leveraging existing techniques are investigated, such as Cloud stream processing, computer networks, and mobile computing.
\end{abstract}


\IEEEpeerreviewmaketitle

\section{Introduction}
The increasingly ubiquitous and powerful smart devices such as sensors and smart phones have been promoting the fast development of data streaming applications, such as augmented reality, interactive gaming, and event monitoring.
The massive data streams produced by these applications have made the Internet of Things (IoT) a major source of big data.
Currently, most mobile and IoT applications adopt the server-client architecture with the frond-end smart devices and the back-end Cloud.
However, the long-distance interactive communications between billions of end devices and the Cloud at the network center would result in two major issues:
 \begin{itemize}
 \item \textbf{Latency.} The end-to-end delay may not meet the requirement of many data streaming applications. For instance, the augmented reality applications typically require a response time of around 10 ms, which is hard to be achieved by using the Could solution with typical end-to-end latency of hundreds of milliseconds.
 \item\textbf{Capacity.} The big data streams may not be affordable by today's network infrastructure. For example, the massive video streams produced by the increasingly deployed cameras put great pressure on today's high-end Metropolitan Area Networks (MANs) with a typical bandwidth of only 100 Gbps¬\cite{satyanarayanan2015edge}.
 \end{itemize}

The emerging Fog architecture¬\cite{chiangfog} paves the way for an ultimate solution that addresses the two issues above, by offloading the back-end computing tasks from the Cloud to Fog servers (i.e. physical or virtual edge servers such as Cisco IOx\footnote{https://developer.cisco.com/site/iox/} and the Cloudlet\footnote{https://en.wikipedia.org/wiki/Cloudlet}) at the network edge.
Due to its shorter distance to the end devices and users, the Fog paradigm has a great potential to not only reduce the backbone Internet traffic, but also to provide services with lower latency and better resilience than the traditional Could paradigm, and therefore are receiving increasing interests from both academia and industry (e.g. the OpenFog Consortium¬\footnote{https://www.openfogconsortium.org}).

This article presents a systemic study of data stream processing and analytics in the context of Fog architecture. Based on the discussions of several typical applications, we present the functional architecture and general models for Fog streaming systems, including the life cycle of data streams, work flow of stream processing tasks, and application-specific processing operations.  
A holistic analysis on the design space of Fog streaming is also presented, with the considerations of key technical issues in four essential dimensions: system, data, human, and optimization.

\section{Fog Streaming Applications}
\begin{figure*}
  \centering
    \includegraphics[width=0.65\textwidth] {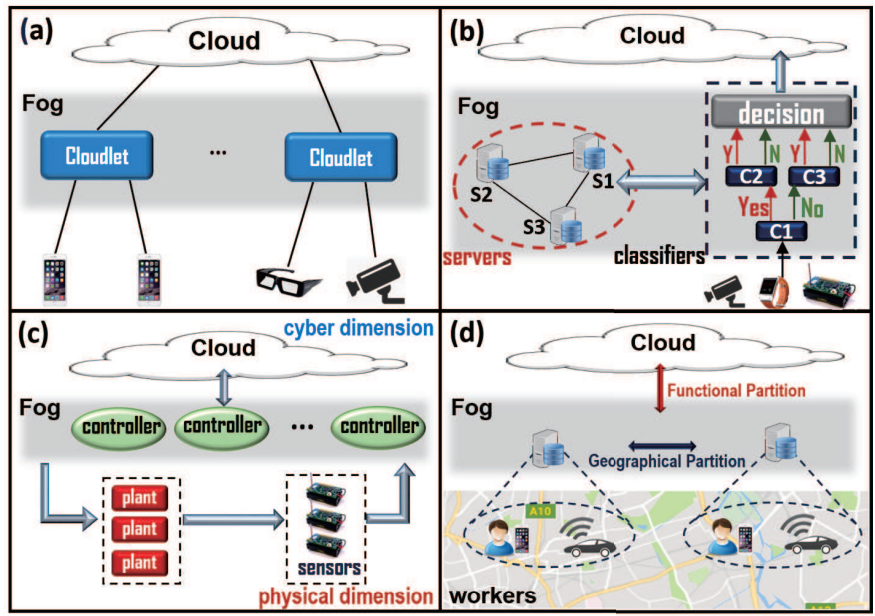}
  \caption{Examples of typical Fog  data streaming Applications.(a) IoT stream query and analytics, (b) Real-time event monitoring, (c)  Networked Control Systems (NCS) for Industrial automation, (d) Real-time Mobile Crowdsensing (MCS).}
  \label{fig:application}
\end{figure*}

This section presents an overview of four typical Fog streaming applications shown in Fig. \ref{fig:application}, in order to demonstrate their typical features, and to clearly illustrate the conceptual Fog architecture in the contexts of different real examples.

\subsection{IoT Stream Query and Analytics}

The fast development of IoT promotes a large class of applications for the high-level query and analytics over the massive sensor data streams.
A typical example of such applications using Fog architecture is Gigasight¬\cite{satyanarayanan2015edge} shown in Fig.\ref{fig:application}(a), an Internet-scale repository system of crowdsoured video streams generated by various cameras, which aims to avoid massive video stream transmissions over the backbone Internet. 
Here, video-processing tasks such as categorization and segmentation are carried out at a Virtual Machine (VM)-based Couldlet over all video streams within the associated Metropolitan Area Network (MAN), and only the video metadata is transmitted to the Cloud for the Internet-wide SQL search on catalog.

Besides Gigasight that explicitly exploits the Internet edge, the existing database systems developed for Wireless Sensor Networks (WSNs)¬\cite{diallo2015distributed} such as TinyDB\footnote{http://telegraph.cs.berkeley.edu/tinydb/overview.html}, implicity adopt the Fog architecture, because both the low-power sensors and the resource-rich gateways (at the network edge) jointly  manage and process sensor data streams. These WSN databases mainly focus on the energy minimization of low-power sensors, and can only provide basic support of sensor data management and SQL-like stream queries.
In addtion, there are several databases such as MongoDB\footnote{https://www.mongodb.com/} for high-performance and NoSQL IoT streaming applications, which can be implemented on both the Cloud servers at the Internet center and the Fog servers at the Internet edge.

\subsection{Real-time Event Monitoring}

Event detection applications such as the vandalism and accident detections are based on the real-time mining of the IoT data streams, which are spatial and temporal correlated in nature.
Fig.\ref{fig:application}(b) illustrates an event detection system using Fog architecture~\cite{canzian2015real}. In this system, the high-level event detection job is divided into different low-level classification tasks (i.e. classifiers), according to the specific application logic and data stream features. The work flow of the event detection job is modelled as a reversed binary tree topology with the root as the data stream source (i.e. sensors), each leaf as an detection result and corresponding actions, and all other vertices as classifiers. These classifiers are allocated to the different Fog servers in a distributed way, by considering the available computing resources of these servers such as CPU, Memory, storage, and network bandwidth.

 \begin{figure*}
  \centering
    \includegraphics[ width=0.75\textwidth] {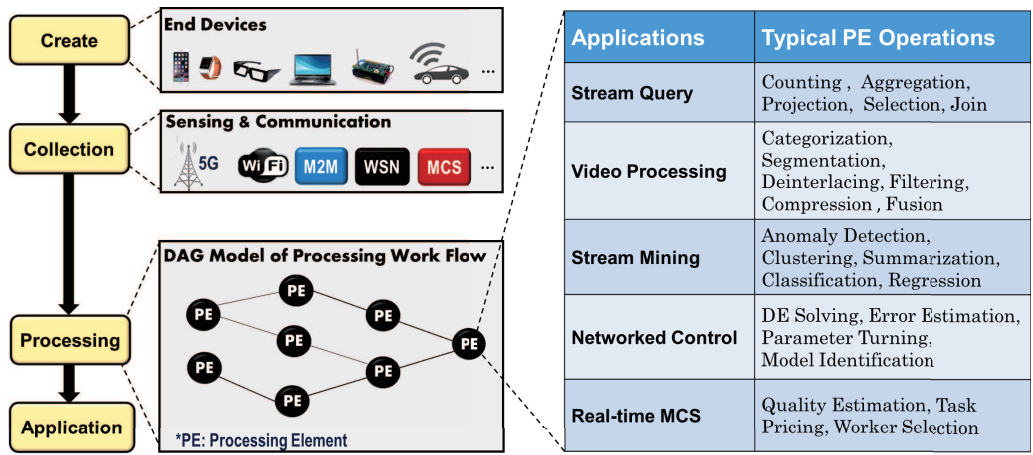}
  \caption{The typical life cycle, DAG model of stream processing work flow, and operations for Fog stream processing tasks.}
  \label{fig:lifecycle}
\end{figure*}

\subsection{Networked Control Systems for Industrial Automation}
As a typical Cyber-Physical System (CPS), the Networked Control System (NCS)~\cite{gupta2010networked} greatly promotes many critical industrial automation applications.
As shown in Fig.\ref{fig:application}(c), the NCS control loop includes controllers, sensors, and control plants (actuators and physical processes), which produce real-time information streams including continuous sensor data flows and control signals, over a communication network.
Adopting the Fog architecture to process such information streams can provide:
\begin{itemize}
\item \textbf{High-quality Communications.} To ensure the desired control performance such as system stability, NCS applications typically require very high-quality communications for the control feedback loop, such as  a 10 ms delay, a 5 Mbps data rate, and a $10^{-8}$ bit error. To satisfy such stringent requirements, local Fog networks should be adopts to minimize distance between all control components, while the Could can provide Internet-scale remote administration services, shown in Fig.\ref{fig:application}(c).

\item \textbf{Rich Computing Resources.} Many advanced NCS applications require computation-intensive control algorithms for solving high-order differential equations, learning system dynamics, and addressing the disturbance and faulty caused by communication uncertainty. Fog servers can provide rich computing resources for these complex control tasks, which cannot be supported by the embedded controllers hosted in the resource-limited end devices.
\end{itemize}

\subsection{Real-time Mobile Crowdsensing}
Mobile Crowdsensing (MCS) is becoming a vital sensing paradigm for urban IoTs, which collects the spatio-temporal sensing contents from enormous participating mobile devices at a city-wide scale\footnote{https://en.wikipedia.org/wiki/Mobile\_Crowdsensing}.
Many MCS applications requires real-time data collection and processing, such as traffic monitoring and collaborative people searching.
In the context of MCS with "human-in-the-loop", the concept of stream processing indicates

\begin{enumerate}
  \item Processing of sensor data flows such as  query and mining, similar to systems with pure machines.
  \item Processing of human-related information streams, such as streams related to incentivization, worker selection, and quality control.
\end{enumerate}

In general, the processing of data streams requires lower latency and higher bandwidth than that of human-related information streams, such as making payment to the participating workers.
As shown in Fig.\ref{fig:application}(d), the hierarchial Fog architecture 
can provide MCS applications with both \textit{geographical partition} of mobile participants and \textit{functional partition} of  different stream processing tasks, resulting in much better performance than current Cloud-based MCS, in terms of scalability, interactive responsive, and bandwidth savings.

\section{Models and Architecture}
This section will present the general models and architecture to characterize the common features of typical Fog streaming systems and applications, including the four examples discussed above.

\subsection{Life Cycle of Fog Data Streams}

As shown in Fig. \ref{fig:lifecycle}, the typical life cycle of the Fog data stream can be divided into the following four stages:
 \begin{enumerate}
\item \textbf{Create.} Fog data streams are majorally created by end  devices, including smart phones, sensors, vehicles, microphones, video cameras, wearable devices, control plants, etc.  It can be seen that the Fog data sources are a subset of Cloud data sources, which also include Internet data produced by social media, logs, emails, financial transactions, databases, E-commerce, and web services. 

\item\textbf{Collection.} At this stage, the created data streams are transmitted from end devices to the Fog servers. A large set of sensing and communication techniques can be utilized for the data collection, including WiFi, 5G cellular networks, WSNs, MCS, Machine-to-machine (M2M) communications. Besides above IoT data collection methods, Cloud data can also be collected using more "soft sensing" methods, such as web crawler for obtaining the web contents.
\item \textbf{Processing.} This stage carries out application-specific  processing tasks based on the collected  data streams at a single fog server, multiple individual fog servers, a small cluster of Fog servers, or a combination of Fog and Cloud servers. Here, some processing tasks are specific for Fog stream applications, such as the networked control and real-time MCS tasks shown in Fig. \ref{fig:lifecycle}.

\item\textbf{Application}. The processing results are consumed by applications and may also be stored for offline batch processing.
 \end{enumerate}

\begin{figure*}
  \centering
    \includegraphics[ width=0.65\textwidth] {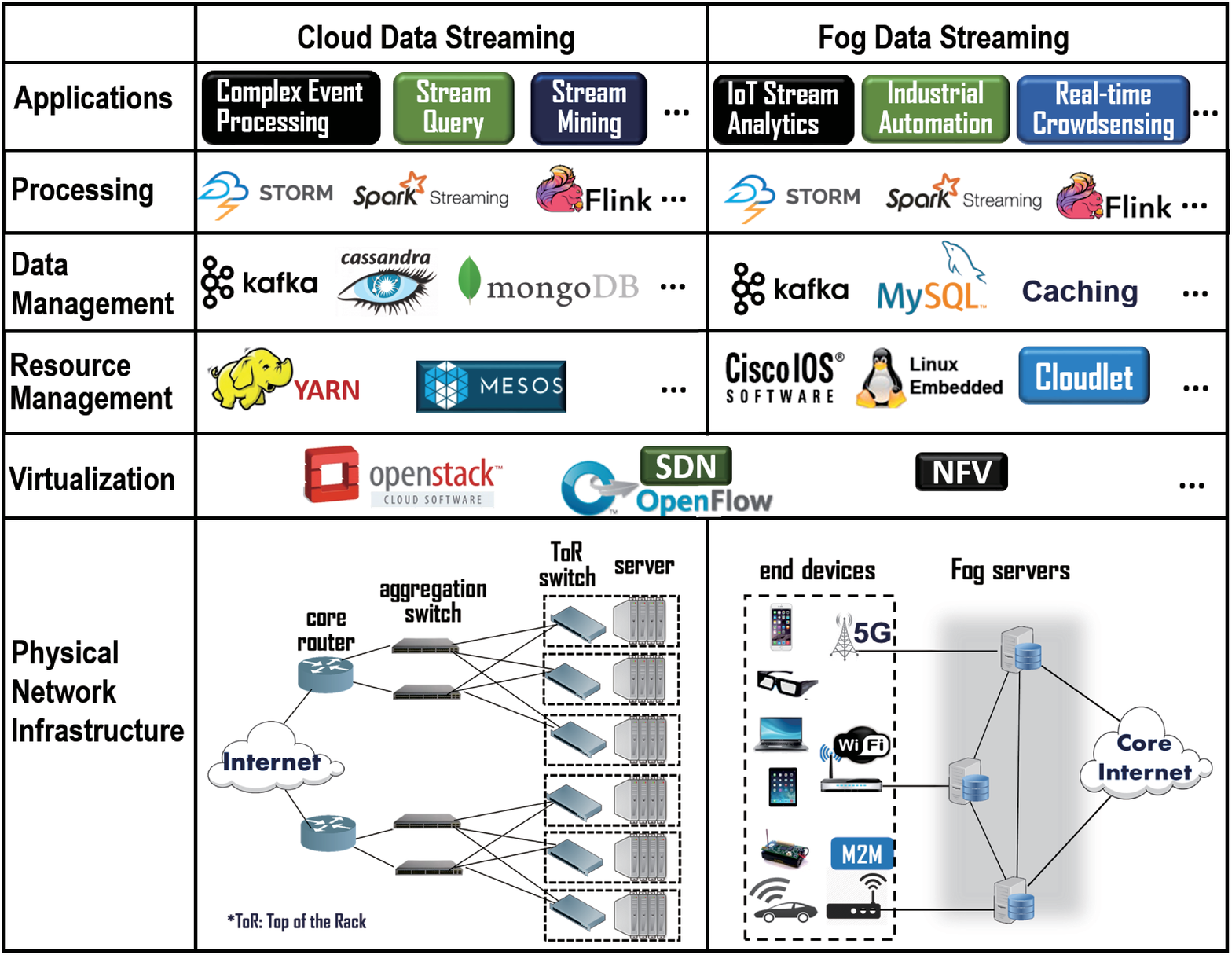}
  \caption{Functional architectures of Cloud and Fog data streaming systems.}
  \label{fig:architecture}
\end{figure*}

It is worthy noting that applications may also produce data streams (e.g. control signals in NCS), resulting in loops in the typical life cycle shown in Fig. \ref{fig:lifecycle}.

\subsection{Work flow and Operations of Fog Stream Processing}
Generally, the high-level logic work flow of a stream processing job can be modelled as
a Directed Acyclic Graph (DAG), as shown in Fig. \ref{fig:lifecycle}. Here,  each vertex represents a Processing Element (PE) performing a variety of low-level computation tasks according to specific Fog streaming applications summarized in Fig. \ref{fig:lifecycle}, and an edge indicates a stream flowing downstream from the producing vertex to the consuming vertex.
For instance, the aforementioned Gigasight application¬\cite{satyanarayanan2015edge} adopts a multistage pipeline for its work flow of denaturing video streams, and the processing process of the event detection application¬\cite{canzian2015real}~follows an binary tree like topology. Both of them are specific forms of DAGs.
Besides naturally describing the high-level abstractions of processing jobs, DAG models greatly facilitate parallel computations of PEs, which are adopted by many high-performance distributed stream processing engines. For example, Apache Storm¬\footnote{http://storm.apache.org/} uses a DAG topology consisting of "spouts" and "bolts", where spouts produce new streams, and bolt consumes injected streams as input and produces streams as output.

\subsection{Fog Data Streaming Architecture}
By using the relatively sophisticate Cloud streaming system as a reference, we propose a Fog streaming architecture shown in Fig.  \ref{fig:architecture}, which includes six functional layers:
 \begin{itemize}
\item \textbf{Application Layer} defines the objective and logic of Fog streaming jobs.
\item \textbf{Processing Layer} carries out the application-specific processing jobs. Recently, a number of real-time stream processing engines have been developed¬\cite{zhang2015memory}, such as Apache Storm,  Spark Streaming¬\footnote{http://spark.apache.org/streaming/}, and Flink¬\footnote{https://flink.apache.org/}. Although these stream processing engines are originally designed for the Cloud and large-scale data centers, they also support the installation on a single or a small cluster of Fog servers. Related technique issues will be discussed in next section.
\item\textbf{Data Management layer} addresses data storage and organization, including file systems, databases, data caches, data warehouses, and data lakes, etc. There are many data management systems working together with stream processing engines in the Cloud, such as the publish-subscribe massaging system Apache Kafka¬\footnote{http://kafka.apache.org/} and the NoSQL database Apache Cassandra¬\footnote{http://cassandra.apache.org/}.  Similar as stream processing engines, these data management system can be applied in Fog servers.
    In addition, data management schemes for local networks such as data-centric caching¬\cite{zhang2013caching} and WSN databases¬\cite{diallo2015distributed} can also be exploited for the Fog data management.
 \item \textbf{Resource Management layer} mainly focuses on the utilization and scheduling of the virtualized system resources, including network and disk I/O bandwidths, CPUs, GPUs, memory, storage, and also energy (e.g. for battery-powered and energy-harvesting devices¬\cite{Shuseninfocom16}).
 \item \textbf{Virtualization Layer} addresses  the configuration and virtualization of the system hardware resources. Virtualization techniques such as Openstack\footnote{https://www.openstack.org/}, Software-Defined Networking (SDN)\footnote{https://en.wikipedia.org/wiki/Software-defined\_networking} and  Network Function Virtualization (NFV)\footnote{https://en.wikipedia.org/wiki/Network\_function\_virtualization} supports both Could and Fog architectures¬\cite{chiangfog}. For instance, both SDN and NFV are considered as the key techniques to facilitate the managements of future 5G networks and the next-generation Internet. Also, a set of Cloudlet-specific APIs are provided in the extension of the Openstack.
     Besides Could-like service paradigms such as Infrastructure as a Service (IaaS), the Fog virtualization can also provide APIs for sensing, caching, mobility, and control services.
 \item \textbf{Physical Network Layer.} As shown in Fig. \ref{fig:architecture}, the Fog system has a much more heterogeneous and dynamic physical network infrastructure than data center networks for the Cloud, although both of them have similar hierarchal network architectures. 
\end{itemize}

It can be seen that many existing Cloud streaming techniques can be leveraged for the Fog.
However,  Fog streaming systems have many features that are significant different from Cloud data streaming systems, including highly delay-sensitive applications, dynamic physical network infrastructures (majorally caused by user mobility), more types of resources (e.g. sensors, actuators, and wireless connectivity), and potentially unreliable services provided by self-interest self users.
Therefore, Could-based approaches may not be able to directly applied in the Fog streaming systems, and new Fog-specific designs considering above features are highly desired.

\section{The Design Space of Fog Data Streaming}

\begin{figure}
  \centering
    \includegraphics[ width=0.3\textwidth] {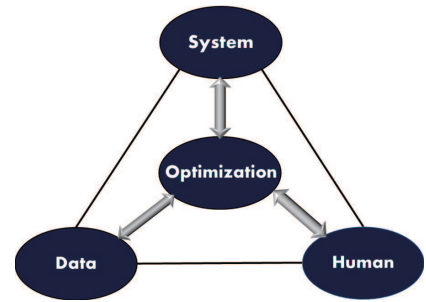}
  \caption{The four dimension in the design space of Fog data streaming.}
  \label{fig:Dimension}
\end{figure}

This section will discuss the design space of Fog data streaming from the viewpoints of four essential dimensions:
system, data, optimization, and Human, where both new design challenges and the issues that arise from applying existing techniques in  Fog streaming will be considered. As shown in Fig. \ref{fig:Dimension}, these four dimensions are not orthogonal, meaning that a technical issue in one (the most relevant) dimension is normally also related to other dimensions.

\subsection{System}
The system dimension refers to the functional components (includes algorithms, protocols, and softwares) related to Fog streaming architecture illustrated in Fig. \ref{fig:architecture}. Specifically,
the following three issues are most critical for establishing the Fog-specific data streaming system.

 \subsubsection{Stream Processing Engine}
  Since there are several well-developed open source stream processing engines such as Apache Storm and Spark Streaming that can run on Fog servers, these is no need to develop a complete new tool for the Fog. However, the following two issues need to be addressed:
 \begin{itemize}
  \item \textbf{Latency-Oriented Processing.} A key objective of existing stream processing engines is to achieve both high throughput and low latency (typical around 100 ms),  while Fog servers typically require much less processing capacity (due to the geographic partition of data stream sources),  but probably more stringent end-to-end delay (less than 10 ms) such as industrial control applications. Therefore, we need to study how to optimize the models and configurations of existing stream processing tools (e.g.  the number of bolts in Storm DAG topology, and mico-batch size of Spark streaming), to support ultra low-latency Fog streaming applications. Bandwidth-hungry Fog streams without such stringent delay requirement (e.g. video streams in Gigasight) can be processed in the same way as normal big data streaming applications or even using offline batch processing tools, but at the Fog servers rather than the Cloud.
\item\textbf{APIs Specific for Fog Streaming.} There exist many libraries that provide rich APIs for advanced data processing such as Apache Mahout\footnote{http://mahout.apache.org/} for machine learning and Spark GraphX\footnote{http://spark.apache.org/graphx/} for graph processing, which can also be used for related Fog streaming applications. However, APIs for some important Fog streaming applications are missing, such as Differential Equation (DE) solvers and control error estimators for Fog-based real-time networked control applications.
 \end{itemize}

 \subsubsection{Streaming Task Partitioning} Due to the three-tier hierarchy of Fog architectures and the geographically distributed end devices and users, the following two types of partitioning of Fog streaming tasks should be considered in the system deign:
 \begin{itemize}
  \item \textbf{Application partitioning} allocates different streaming tasks to end devices, Fog servers, and Could servers, according to available resources, privacy concerns, latency requirement, fault-tolerance, etc.  Different granularity levels of partitioning can be adopted such as partitioning of multiple applications,  multiple data streams, and functional components in a single application. Existing work \cite{yang2013framework} for the Mobile Cloud Computing (MCC) paradigm (typically two-tierd architecture with end devices and Cloud servers) can also be extended for the three-tier Fog.
  \item \textbf{Geographic partitioning} allocates streaming tasks among different Fog servers. Here, load balancing among Fog servers is particular important for Fog systems with heterogeneous server capacities and end devices density.
 \end{itemize}

\subsubsection{Streaming Service Migration}
When a user moves away from the Fog server that he or she is currently using, the corresponding streaming service should be
migrated to a new server seamlessly, with the minimal degradation of end-to-end streaming quality.
However, existing approaches for Could computing (e.g. the Live Migration¬\footnote{https:\/\/en.wikipedia.org\/wiki\/Live\_migration})
would perform poorly in the Fog environment due to the high uncertainty and dynamics caused by user mobility, while the current Fog service migration schemes such as \cite{wang2015dynamic} are limited to unrealistic mobility patterns.  Therefore, the design of new streaming-specific service migration algorithms with
real-time and faulty-tolerance supports are highly desired.

\subsection{Data}
Existing data streaming algorithms (including data stream acquisition and mining) assume that their underlying computing infrastructure is a single server, a local distributed network (e.g. a WSN), or the Cloud.  The Fog paradigm creates several new design opportunities for these algorithms.

\subsubsection{Data Stream Acquisition.} Data stream acquisition refers to the processes of sensing and data collection from local networks to the Fog servers.
\begin{itemize}
 \item \textbf{Sensing.} The spatio-temporal correlation of IoT data streams enables advanced sensing techniques such as compressive sensing to minimize the sampling rate and therefore the network traffic loads.
      However, for citywide-scale (or real-time multimedia streaming) applications¬\cite{wang2015ccs}, current compressive sensing algorithms would suffer from heavy computations for the sensing matrix reconstruction, and intensive communications between end devices and the Could server.
      It is promising to address these issues by exploiting the three-tier Fog architecture.
      For instance, each Fog server communicated with its associated end devices, and reconstructs a local sensing sub-matrix, based on which the Cloud server can further recover the global one.
      To achieve this, new compressive sensing algorithms utilizing the hierarchical Fog architecture should be designed.

 \item \textbf{Data Stream Cleansing.}
     To improve data acquisition quality, raw data streams should be processed by removing the abnormal (faulty, incorrect, or false) data records. Many real-time anomaly detection algorithms such as \cite{chen2015TIE} are based on exploiting spatio-temporal correlations of the raw data time series.
      Since data sources in geographical proximity are more likely to be correlated, each Fog server can perform as the local processing center to detect anomalies of the highly-correlated data streams collected from its associated local network. This results in much higher fault-tolerance than using end devices to perform the detection tasks \cite{chen2015TIE}.
\end{itemize}

\subsubsection{Stream Mining and Analyltics}
Existing research on real-time data mining such as \cite{canzian2015real} provide the theoretical foundation of  distributed stream mining (e.g. feature abstraction and classification) in networked systems, such as Fog systems. In addition, the recently released open source software (e.g. TensorFlow¬\footnote{https://www.tensorflow.org/}) significantly facilitate the implementation of advanced machine learning and data mining algorithms (e.g. deep neural networks) in the Fog servers and even end devices (e.g. Mobile TensorFlow). Although these the theoretical results and engineering supports open a new door for Fog streaming mining and analytics, a set of new challenges arise, especially how to balance the computation loads among Fog servers and end devices at real-time, while ensuring the mining performance.

\subsection{Human}
Compared with the Cloud,  Fog systems are closer to the users and end devices (thus their owners). Therefore, humans play a more important role and their behaviors must be considered in the holistic Fog streaming design.

\subsubsection{Pricing and Icentivization} From economics viewpoint, people in the Fog streaming system can be classified into two types:
\begin{itemize}
\item \textbf{Service Providers} including private owners of end devices and Fog servers, who have data and resource, and can provide various streaming services.
\item\textbf{Service Consumers} who discover, subscribe, and consume the streaming services.
\end{itemize}
Due to the inherent self-interest and strategic behaviors of both service providers and consumers, proper incentivization and pricing mechanisms are essential to ensure the efficiency and trustworthiness of economic activities between service providers and consumers.
There exists a large body of related research such as cloud data pricing, smart grid pricing, and crowdsensing auctions. These results can be leveraged for Fog streaming applications, while specific attentions should be taken on addressing the heterogeneity and dynamics of Fog systems.

\subsubsection{Privacy} An important issue in Fog streaming is to balance the tradeoff between the data value and the risk of privacy exposure.
For instance, Gigasight¬\cite{satyanarayanan2015edge} performs the denaturing process of video streams at the network-edge Cloudlet to abstract video features while preserving privacy of video providers.
Actually, the hierarchical Fog architecture can be exploited to provide resilient privacy preservations at each of the three tiers, according to different application contexts.

\subsubsection{Quality Control}
Since crowdsoured workers are different in their problem solving abilities, quality control is essential for the real-time mobile crowdsensing, a typical Fog streaming application mentioned before.
For instance, in the real-time speech captioning application¬\cite{lasecki2013warping}, audio streams with different speaking rates are allocated to the crowdsourced workers according to their abilities, to ensure their online task completion qualities.
With the Fog architecture, both the functional partitioning of task allocations and the geographic partitioning of crowdsourced workers can be exploited to optimize the real-time quality control process.

\begin{figure}
  \centering
    \includegraphics[ width=0.45\textwidth] {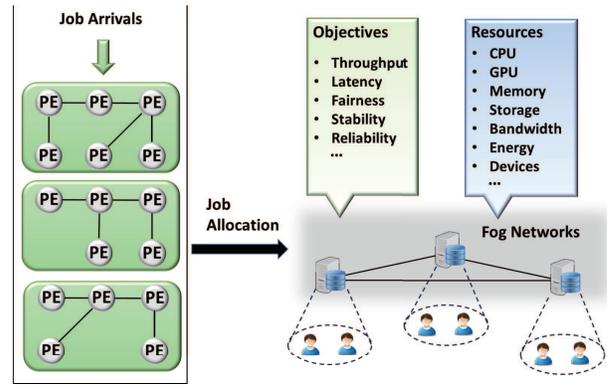}
  \caption{Dynamic resource allocation for DAG-like Fog streaming jobs.}
  \label{fig:taskallocation}
\end{figure}

\begin{figure*}
  \centering
    \includegraphics[ width=0.6\textwidth] {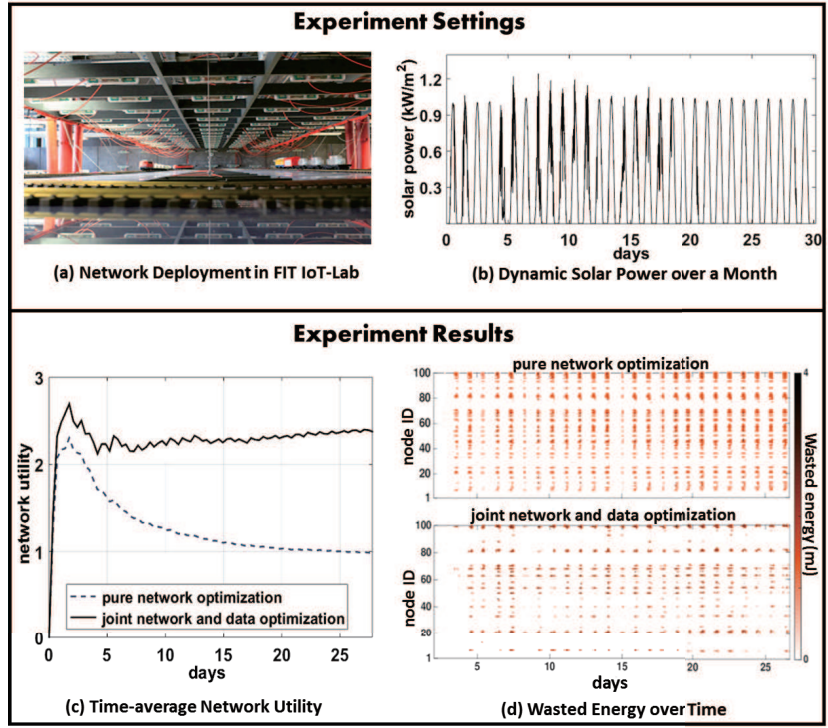}
  \caption{In-network data processing~\cite{Shuseninfocom16} as an example to demonstrate the performance gain of joint data and network optimization.}
  \label{fig:harvesting}
\end{figure*}

\subsection{Optimization}
To better understand and solve the issues in system, data, and human dimensions, new theoretical models and methods are required, which are referred as the optimization dimension.

\subsubsection{Dynamic Optimization.}  Fog streaming systems are inherently dynamic and uncertain, caused by various reasons, including mobility, wireless communications, physical events, unreliable data providers, and faulty-prone sensors, and server failures, etc.
Therefore, analytical models of Fog streaming problems should take a specific attention on  corresponding dynamics and uncertainties. For instance, the algorithm proposed in \cite{wang2015dynamic}  uses Markov decision process to address the edge-cloud service migration caused by mobility, and
algorithm proposed in \cite{edgecouldmobihoc2016} can support dynamic service configurations with arbitrary stochastic processes of service arrivals.

\subsubsection{Complex Resource Allocation}
All stream processing jobs consume resources.
As shown in Fig. \ref{fig:taskallocation}, due to the complex DAG structure of dynamically arrived streaming jobs and the heterogeneous types of resources in the networked fog system, resource allocation for Fog data streaming are challenging optimization problems.
Ghaderi \textit{et al}¬\cite{ghaderi2016scheduling} propose an optimization approach for the resource allocation of DAG-like streaming jobs of Apache Storm for the data center networks, which shares similar topology as Fog network infrastructure shown in Fig. \ref{fig:architecture}, and therefore is possible to be extended to support Fog stream processing.

\subsubsection{Optimization over System, Data, and Human Dimensions}
Due to the multidisciplinary nature of Fog data streaming, joint optimization over the system, data, and human dimensions would outperform the optimization in each individual dimension.
For instance, our in-network processing algorithm¬\cite{Shuseninfocom16} that optimizes the data processing and network (system) operations jointly manages to achieve better practical performance than pure network optimization, in terms of energy resource utilization and network throughput, as shown in Fig. \ref{fig:harvesting}.
To achieve cross-dimension optimization, new analytical models and methods should be developed by leveraging  mathematical methods in each dimension, such as queuing theory for system, signal processing for data, and game theory for the human dimension.

\section{Conclusion}

This article presents a systemic investigation on data stream processing and analytics in the context of Fog architecture.
We study four typical Fog streaming applications, including IoT stream analytics, event monitoring, networked control, and real-time mobile crowdsourcing, which demonstrate their common properties and the multi-disciplinary nature of Fog streaming research.
These practical applications result in the discussions on the general Fog streaming models and architecture, as well as the opportunities and challenges in the future design, in terms of networked systems, data processing and management, human factors, and optimization methods.
We  expect that the increasingly important roles of both network edge and stream processing will further promote their combinations, and thus the development of Fog data streaming in both academia and industry.

\section*{ACKNOWLEDGMENTS}
This work is sponsored by  China "1000
Young Talents Program" and "Young Talent Support Plan" of Xi'an
Jiaotong University.

\end{document}